\documentclass[reprint,superscriptaddress,nofootinbib,onecolumn,aps,pra,longbibliography]{revtex4-2}

\usepackage[utf8]{inputenc}

\usepackage{amsmath}
\usepackage{amssymb}
\usepackage{amsfonts}
\usepackage{bm}
\usepackage{graphicx}
\usepackage{xcolor}
\usepackage{hyperref}
\usepackage{enumitem}
\usepackage[normalem]{ulem}

\usepackage[caption=false]{subfig}
\usepackage{graphicx}

\usepackage{fancyhdr}
\DeclareMathOperator{\sech}{sech}
\newcommand{\p}{\partial}

\newcommand{\e}{\bm{\hat{e}}}

\newcommand{\Spin}{\bm{S}}
\newcommand{\magn}{\bm{m}}
\newcommand{\nagn}{\bm{n}}
\newcommand{\Heff}{\bm{F}}
\newcommand{\heff}{\bm{f}}

\newcommand{\DM}{D}
\newcommand{\dm}{\lambda}
\newcommand{\Anisotropy}{K}
\newcommand{\storque}{\beta}
\newcommand{\ptorque}{\bm{\hat{p}}}

\newcommand{\losc}{\ell}
\newcommand{\Je}{J_e}

\newcommand{\Energy}{E}

\newcommand{\tmagn}{{\mathcal{M}}}

\begin{document}

\title{Antiferromagnetic domain walls under spin-orbit torque}
\author{G. Theodorou}
\affiliation{Department of Mathematics and Applied Mathematics, University of Crete, 70013 Heraklion, Crete, Greece}
\affiliation{Institute of Applied and Computational Mathematics, FORTH, Heraklion, Crete, Greece}
\author{S. Komineas}
\affiliation{Department of Mathematics and Applied Mathematics, University of Crete, 70013 Heraklion, Crete, Greece}
\affiliation{Institute of Applied and Computational Mathematics, FORTH, Heraklion, Crete, Greece}
\date{\today}

\begin{abstract}
Domain walls in antiferromagnets under a spin-polarized current present dynamical behavior that is not observed in ferromagnets, and it is tunable by the current polarization.
Precessional dynamics is obtained for perpendicular spin polarization.
In-plane polarization gives propagating walls.
We obtain the velocity as a function of current by a perturbation method for low velocities, and the wall profile is found to lack a definite parity.
For high velocities, a power-law decay develops in the trailing tail of the wall.
The main features of the wall profile are manifest in a direct solution of an equation that is valid in a limiting case.
Oscillatory motion of domain walls is obtained for a spin polarization that has both perpendicular and in-plane components, and an analytical description is given.
We discuss the modifications of the dynamics when a Dzyaloshinskii-Moriya interaction is present.
Finally, we give the magnetization of the dynamical walls and find that this can become large, providing a potential method for observations.
\end{abstract}

\let\clearpage\relax
\maketitle

\section{Introduction}
\label{sec:introduction}

Antiferromagnetic domain walls have attracted attention since decades ago \cite{1990_PR_KosevichIvanovKovalev,1991_AdvPhys_MikeskaSteiner} and their study has led to remarkable conclusions \cite{1983_PRL_Haldane}.
Interest has grown in recent years as antiferromagnetic order can now be observed in a more efficient way and domain walls can be directly imaged 
\cite{2020_npjQM_CheongFiebig,2021_NatPhys_HedrichShekaMakarov,2022_acsnano_GrigorevKlaeuiDemsar,2022_PRB_SatoArima}.
In addition, experimental techniques allow for optical
\cite{2023_advMater_MeerKlaeui} and electrical
\cite{2018_NatNano_WadleyJungwidth,2022_PRB_SatoArima} manipulation of domain walls, while current-induced switching can be achieved by domain wall motion \cite{2019_PRL_BaldratiKlaeui}.
The efficient manipulation of antiferromagnets (AFMs) has the potential to lead to significant advantages over the widely used ferromagnets, as antiferromagnetic order produces no stray fields thus making it easier to control and also robust under moderate external fields.

The dynamics in antiferromagnets is described by extensions of the nonlinear $\sigma$-model for the N\'eel vector \cite{1990_PR_KosevichIvanovKovalev,1991_AdvPhys_MikeskaSteiner} and is radically different from the dynamics of magnetization in ferromagnets, potentially leading to advantages.
It allows for a wider range of dynamical behaviors including Newtonian and relativistic dynamics \cite{1995_PRB_Papanicolaou,1995_PRL_IvanovKolezhuk,2018_LTP_GalkinaIvanov,2020_SciPost_KomineasPapanicolaou} as the $\sigma$-model is second order in time.
Dynamics in AFM can be probed by spin-polarized currents, and antiferromagnetic spintronics is now being developed in a fashion analogous to ferromagnets \cite{2016_NatNano_Jungwirth,2024_npjSpin_DalDinWadley}.
Spin-orbit torques drive antiferromagnetic domain walls faster than ferromagnetic domain walls
\cite{2016_PRL_ShiinoLee,2016_PRL_GomonayJungwirthSinova,2020_PRB_TejerinaFinocchio}
while they can also be manipulated by spin-transfer torques  \cite{2011_PRB_SwavingDuine,2011_PRL_HalsBrataas,2014_PRL_TvetenBrataas}. 

We study the dynamics of antiferromagnetic domain walls under spin-orbit torque.
The detailed features of propagating domain walls are studied.
We find that the profile of the domain wall is asymmetric and develops a power-law tail for large currents.
Its velocity has a nontrivial dependence on the current for large but also for moderate currents.
Precessional dynamics, analogous to that discussed in the conservative model \cite{1983_PRL_Haldane,1983_SPJETP_BaryakhtarIvanov,1995_PRL_IvanovKolezhuk,2017_LTP_GalkinaOvcharovIvanov}, is spontaneously obtained as a stable steady state.
For a spin current polarization that has both an in-plane and a perpendicular component, the domain wall undergoes oscillating motion between two positions.
This motion has two variants: the same oscillation is obtained for two different wall configurations.
The dynamics of the domain wall under spin torque has been studied using a collective coordinate approach and employing the Lorentz transformation \cite{2016_PRL_ShiinoLee,2022_PRApplied_OvcharovGalkinaIvanov,2023_PRApplied_OvcharovIvanov}.
The present work studies the full model and goes beyond the idealized $\sigma$-model in terms of the methods used and the results obtained.

We discuss the magnetic moment carried by the studied antiferromagnetic textures.
This is of particular importance, as it could allow for the observation of textures and gives a handle to make them functional.
For domain walls, a net magnetic moment has been theoretically predicted within a strictly one-dimensional model \cite{1995_PRB_Papanicolaou,1998_JPCM_Papanicolaou},
and it has subsequently been observed \cite{2006_NatMater_Bode,2017_SciRep_Hirose}.
For a dynamical wall, the time dependence of the N\'eel order parameter gives rise to an additional magnetization. 

The paper is organized as follows.
Sec.~\ref{sec:model} introduces the discrete and continuum models.
Sec.~\ref{sec:propagatingDW} is a study of propagating domain walls.
Sec.~\ref{sec:precession} is a study of the precessional dynamics of domain walls.
Sec.~\ref{sec:oscillations} studies oscillatory motion of domain walls.
Sec.~\ref{sec:conclusions} contains our concluding remarks. The data supporting the findings in this study is available via Zenodo at https://doi.org/10.5281/zenodo.17898266.

\section{The model}
\label{sec:model}

We consider a spin chain $\Spin_i$ with antiferromagnetic exchange, Dzyaloshinskii-Moriya (DM) interaction, and perpendicular anisotropy.
The magnetic energy is
\begin{equation} \label{eq:energyDiscrete}
    \Energy^d = \sum_i J\, \Spin_i\cdot \Spin_{i+1}
     + \DM\,\e_2\cdot (\Spin_i\times\Spin_{i+1})
    + \frac{\Anisotropy}{2}\, [s^2 - (\Spin_i)_3^2],
\end{equation}
where $J, \DM, \Anisotropy$ are positive parameters and the spin variables have length $|\Spin_i|=s$ with $s$ a dimensionless number.
The energy $\Energy^d$ has units of inverse time.
We will study antiferromagnetic domain walls and their dynamics.
We will mainly study the case with no DM interaction ($\DM=0$), but will comment on the effect of DM in all the studied cases.
For probing the dynamics of the antiferromagnet, we will assume spin torque produced by polarized electrons injected into the sample.
The equations of motion for the spins are 
\begin{equation} 
\label{eq:Heisenberg}
 \frac{\p \Spin_i}{\p t} = \Spin_i\times \left( \Heff_i - \tilde{\alpha} \frac{\p \Spin_i}{\p t} - \tilde{\storque}\,\Spin_i\times\ptorque \right),  \qquad \Heff_i = - \frac{\p \Energy^d}{\p\Spin_i} = -J(\Spin_{i-1}+\Spin_{i+1}) + \Anisotropy \Spin_i\e_3
\end{equation}
where $\e_3$ is the unit vector in the third direction, $\tilde{\alpha}$ is the damping parameter, $\ptorque$ is the spin current polarization and
\[
\tilde{\storque}=\frac{\Je A b}{e s^2}
\]
is the spin torque parameter, with $\Je$ the current density, $A$ the area occupied by a single spin, $e$ the electron charge, and $b$ a constant that is related to the degree of spin polarization.

For a theoretical study, a continuum model for the equations of motion is necessary.
This is usually obtained when we consider a dimerization of the spin lattice and define the N\'eel vector $\nagn_\alpha$ and magnetization $\magn_\alpha$ that are the normalized difference and sum of neighboring spins, respectively, at each dimer site $\alpha$.
While we study configurations that vary in one dimension, we assume that the actual spin lattice is two-dimensional, such as a film (this requires considering spins $\Spin_{i,j}$ and generalizing the energy \eqref{eq:energyDiscrete} accordingly).
Instead of dimers, we consider groups of four neighboring spins on the vertices of a square (termed ``tetramers'') since they comply with the symmetry of a two-dimensional square spin lattice \cite{1998_NL_KomineasPapanicolaou}.
The distance between neighboring tetramer sites is defined to be $2\epsilon$ where
\[
\epsilon = \sqrt{\frac{\Anisotropy}{J}}
\]
is a small parameter.
In the limit $\epsilon\to 0$, a continuous N\'eel vector field $\nagn=\nagn(x,\tau)$ is obtained with $|\nagn|=1$, where $x$ and $\tau$ are scaled space and time variables.
The equation of motion in the continuum is an extension of the nonlinear $\sigma$-model for the N\'eel vector $\nagn$ with the addition of damping and a spin torque term \cite{1979_SJLTP_BaryakhtarIvanov,1991_AdvPhys_MikeskaSteiner,2010_PRB_GomonayLoktev},
\begin{equation} \label{eq:sigmaModel_1D_spinTorque}
        \nagn\times (\ddot{\nagn} - \heff + \storque\nagn\times\ptorque + \alpha\dot{\nagn} ) = 0,\qquad \heff = \nagn'' - 2\dm\,\e_2\times\nagn' + n_3\e_3
\end{equation}
where the dot denotes differentiation in time and the prime denotes differentiation in space.
Actual distances are given by $a x/\epsilon$ where $a$ is the physical distance between neighboring spins.
The scaled time is defined as
\begin{equation}\label{eq:physicalTime}
    \tau = 2\sqrt{2}\,\epsilon s J\,t.
\end{equation}
The scaled parameters are related to the parameters in the discrete model by
\begin{equation} \label{eq:scaledParameters}
\dm = \frac{\DM}{\epsilon J},\qquad
\alpha = \frac{2\sqrt{2}}{\epsilon}s\tilde{\alpha},\qquad \storque = \frac{\tilde{\storque}}{\epsilon^2 J}.
\end{equation}

For an estimation of the model parameters, let us choose $J=10^{-21}\,{\rm Joule}/\hbar \sim 10^{13}\,{\rm sec^{-1}}$ and $\Anisotropy=0.0025 J$, obtaining $\epsilon=0.05$.
If we further choose $A \sim 10^{-20}\,{\rm m^2},\;b\sim 0.1,\; s = 1$, then $\storque \sim J_e/(4\times 10^{12}\,{\rm A/m^2})$, and for $\Je \sim 10^{12}\,{\rm A/m^2}$ it gives $\storque \sim 0.25$.
The unit of length is that of the domain wall width, which could be on the order of $10\,{\rm nm}$.
The unit of time, from Eq.~\eqref{eq:physicalTime} is $1/(2\sqrt{2}\,\epsilon s J) \sim 0.7\,{\rm ps}$.
The unit of velocity is $2\sqrt{2}\, a s J \sim 1.4\times 10^4\,{\rm m/sec}$ (assuming a distance between ions $a=0.5\,{\rm nm})$.

The magnetization is given, when tetramers are used, by \cite{1998_NL_KomineasPapanicolaou}
\begin{equation} \label{eq:magnetization_tetramer}
    \magn = \frac{\epsilon}{2\sqrt{2}}\,\nagn\times\dot{\nagn}.
\end{equation}
If a one-dimensional antiferromagnet is considered (such as a superlattice structure), one should consider dimers and in this case the magnetization has an additional term $(\epsilon/2)\nagn'$ \cite{1991_AdvPhys_MikeskaSteiner,1995_PRB_Papanicolaou}.
In this paper, we will only consider the magnetization in Eq.~\eqref{eq:magnetization_tetramer}.
The physical spin is $\bm{\mu}=4s\,\magn$ at every tetramer site, and the total magnetization $\bm{\tmagn}$ is found by summing over all tetramers.
In the limit $\epsilon\to 0$, we integrate over space \cite{1995_PRB_Papanicolaou} and have the formula
\begin{equation} \label{eq:totalMagnetization}
    \bm{\tmagn} = \frac{4s}{2\epsilon} \int \magn\,dx = \frac{s}{\sqrt{2}} \int \nagn\times\dot{\nagn}\,dx.
\end{equation}

The study starts with the remark that Eq.~\eqref{eq:sigmaModel_1D_spinTorque}, when no spin current is present, has a static solution which is a standard ($180^\circ$) domain wall.
We will study the dynamics of the domain wall under spin current for various cases of spin polarization.

All numerical results that will be presented in this work have been obtained by simulating the spin lattice.
Starting from Eq.~\eqref{eq:Heisenberg}, we obtain the standard Landau-Lifshitz-Gilbert equation for the time derivative of the spin vectors and iterate it in time using a Runge-Kutta method.

We typically use the parameter values $s=1, \Anisotropy/J=0.0025$, that give $\epsilon=0.05$.
Our numerical mesh contains 12000 sites, which gives a space variable in the continuum theory from $x=-300$ to $x=300$.

\section{Propagating walls}
\label{sec:propagatingDW}

We consider a current polarized in the $\ptorque=\e_2$ direction.
We study numerically the behavior of domain walls by solving in time the discrete Eq.~\eqref{eq:Heisenberg} using as an initial condition a static N\'eel type domain wall.
Initially, the domain wall is deformed and accelerated.
Eventually, the system reaches a steady state propagating with constant velocity, as reported in Refs.~\cite{2016_PRL_ShiinoLee,2022_PRApplied_OvcharovGalkinaIvanov}.
In addition, in the case where the simulation starts from a wall that is not perfectly N\'eel, this still reaches the same steady state.
Only when the simulation starts from a perfectly Bloch wall does this remain static, although it is deformed.
Traveling domain walls are obtained with or without the DM term.
In the following, we present results for the case without DM (set $\DM=0$).

\begin{figure}[t]
    \centering
    {\includegraphics[width=7cm]{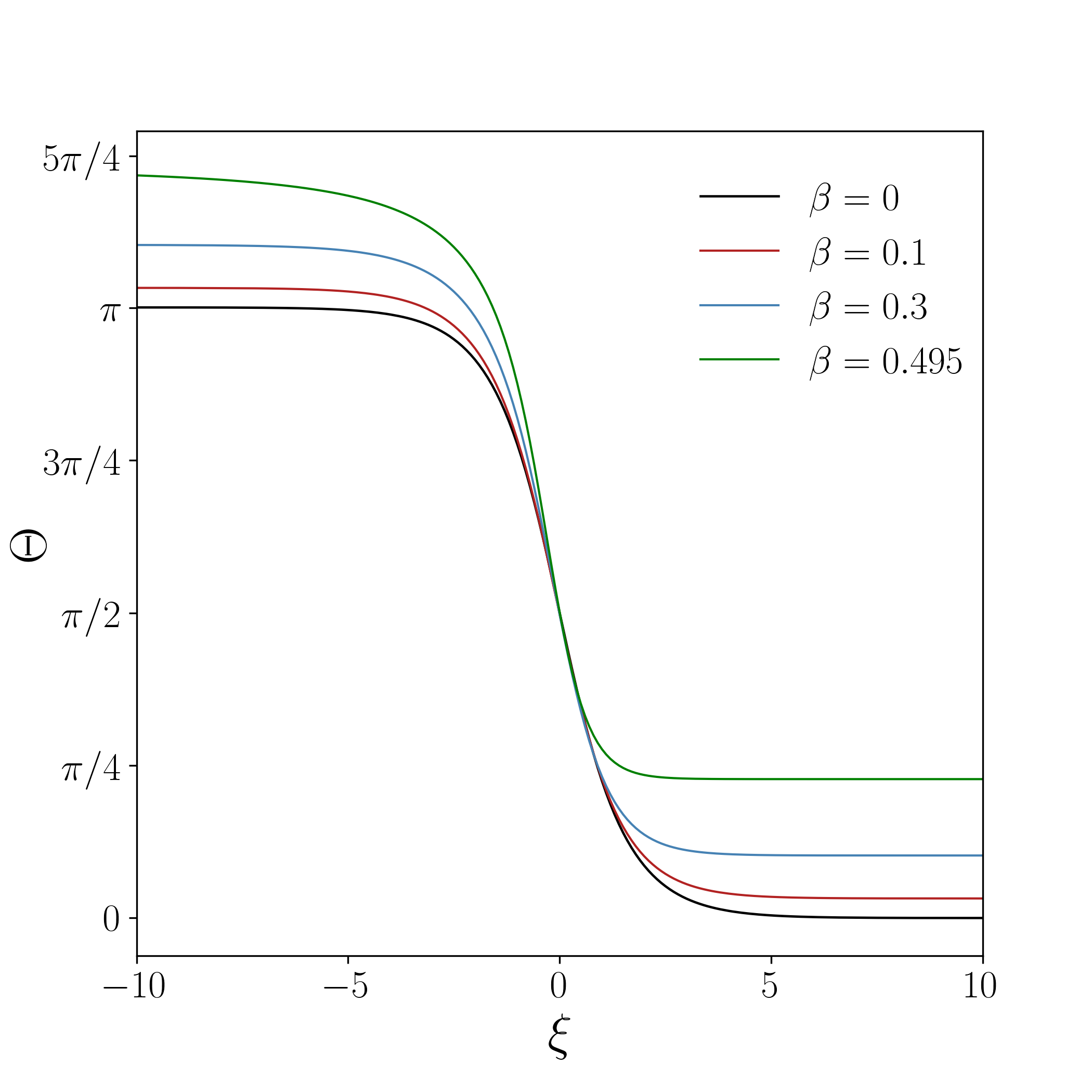}}\qquad
    \caption{Traveling domain walls with a constant velocity found numerically for damping parameter value $a=0.8$ (the large value of $\alpha$ is used to amplify some features of the wall).
    The applied spin torques are $\storque = 0.1$, $\storque= 0.3$ and $\storque= 0.495$ (the domain wall for $\storque=0$ is shown for comparison) and the corresponding velocities are $v\approx 0.19$, $v\approx 0.53$ and $v\approx 0.81$.
    The N\'eel vector at the ends of the wall is tilted with respect to the south ($\Theta=\pi$) and the north pole ($\Theta=0$).
    The wall profile for $v\neq 0$ is not symmetric with respect to the center of the wall.
    We have power-law decay for $\xi\to-\infty$ and exponential decay for $\xi\to\infty$.
    }
    \label{fig:traveling_solution2}
\end{figure}
 
Fig.~\ref{fig:traveling_solution2} shows the results of simulations for steady-state domain walls for four different values of the parameter $\storque$ and for damping $\alpha=0.8$.
We use the spherical parametrization for the N\'eel vector,
\begin{equation} \label{eq:sphericalParametrization}
n_1 = \sin\Theta \cos\Phi,\quad n_2 = \sin\Theta \sin\Phi,\quad n_3 = \cos\Theta
\end{equation}
and $\Phi=0$ for a N\'eel wall.
The N\'eel vector at the ends of the wall is tilted with respect to the south and the north pole.
This feature is apparent in figures in Ref.~\cite{2016_PRL_ShiinoLee}.
The wall profile is not symmetric with respect to the center of the wall, a phenomenon that becomes more evident at higher values of current and velocity.
For a smaller damping $\alpha$, the results are similar, but it takes longer to reach the steady state. 

For the study of translational motion, we assume the traveling wave ansatz $\nagn=\nagn(x-v\tau)$.
The continuum model \eqref{eq:sigmaModel_1D_spinTorque} gives
\begin{subequations} \label{eq:ThetaPhi_traveling_spinTorque}
\begin{align}
& (1-v^2)\Theta''-\left[1 + (1-v^2){\Phi'}^2 \right] \sin \Theta\cos\Theta + \alpha v\Theta' = -\storque\cos\Phi. \label{eq:ThetaPhi_traveling_spinTorque_Theta} \\
    & (1-v^2)\Phi'' \sin\Theta +2(1-v^2) \Theta' \Phi' \cos\Theta + \alpha v\Phi'\sin\Theta = \storque \cos\Theta \sin\Phi\label{eq:ThetaPhi_traveling_spinTorque_Phi}
\end{align}
\end{subequations}
where the prime denotes differentiation with respect to the argument $\xi=x-v\tau$.
Polarized (uniform) states of Eqs.~\eqref{eq:ThetaPhi_traveling_spinTorque} satisfy
\[
\storque \cos\Phi = \sin\Theta \cos\Theta,\qquad
\storque \cos\Theta \sin\Phi = 0.
\]
The second equation is satisfied for $\Phi=0$ or $\pi$ (i.e., $n_2=0$).
It is convenient to set $\Phi=0$ and let $0 \leq \Theta \leq 2\pi$.
The first equation gives \cite{2023_PRApplied_OvcharovIvanov}
\begin{equation} \label{eq:polarized_tilted}
\sin(2\Theta) = 2\storque\qquad \text{for}\qquad
-\frac{1}{2} < \storque < \frac{1}{2}.
\end{equation}
Eq.~\eqref{eq:polarized_tilted} has two pairs of solutions for each value of $\storque$.
For $\storque=0$, the first pair of polarized states are $\Theta=0,\pi$. 
For $\storque\neq0$, these are tilted with respect to the north and the south pole.
Let us call this pair $\Theta_P,\, \pi + \Theta_P$.
The second pair gives polarized states that are on the equator $\Theta=\pi/2, 3\pi/2$ for $\storque=0$ and are tilted close to the equator for $\storque\neq0$.
The latter represent unstable states, as they are almost orthogonal to the easy axis, and they will not be studied further here.

To find nontrivial solutions for Eq.~\eqref{eq:ThetaPhi_traveling_spinTorque}, we note that Eq.~\eqref{eq:ThetaPhi_traveling_spinTorque_Phi} is satisfied for $\Phi=0$ and Eq.~\eqref{eq:ThetaPhi_traveling_spinTorque_Theta} becomes 
\begin{equation} \label{eq:Newton_dampingForcing}
  \left(1-v^2\right) \Theta'' + \alpha v\Theta' = \sin \Theta\cos\Theta - \storque.
\end{equation}
This is of the form of Newton's equation for a particle in a potential including damping and a constant external force.
A perturbed sine-Gordon model that gives Eq.~\eqref{eq:Newton_dampingForcing} upon using the traveling wave ansatz was employed in Ref.~\cite{1978_PRA_MacLaughlinScott} for studying fluxons of magnetic field and their interactions in a Josephson-junction transmission line.
A method was presented for studying the dynamics of solitons of the completely integrable sine-Gordon equation under the perturbation.

Early studies on propagating domain walls in AFM \cite{1983_PRL_Haldane,1983_SPJETP_BaryakhtarIvanov,2018_LTP_GalkinaIvanov} have been based on the Lorentz invariance of the conservative model ($\alpha=0,\,\storque=0$).
We study Eq.~\eqref{eq:Newton_dampingForcing} where Lorentz invariance is broken due to the presence of a damping term, thus going beyond the results of early studies and of more recent work \cite{2016_PRL_ShiinoLee,2016_PRL_GomonayJungwirthSinova,2022_PRApplied_OvcharovGalkinaIvanov,2023_PRApplied_OvcharovIvanov}.

\begin{figure}
    \centering
    \includegraphics[width=6cm]{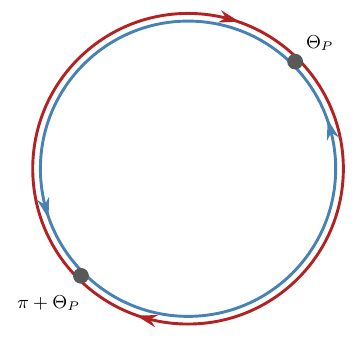}
    \caption{Domain walls are represented by semicircles connecting the polarized states $\Theta=\Theta_P$ and $\Theta=\pi+\Theta_P$.
    The two semicircles in each one circle give two equivalent representations of the same wall.
    That means that each circle represents one different kind of domain wall.}
    \label{fig:DWonCircle}
\end{figure}

For the construction of domain walls, $\Theta$ should approach $\Theta_P$ and $\pi + \Theta_P$ at the two ends of the spin chain.
There are two kinds of domain walls represented on the circles in Fig.~\ref{fig:DWonCircle}.
The first one is constructed by setting $\Theta(-\infty)=\pi+\Theta_P$ and $\Theta(\infty)=\Theta_P$ and is represented by a semicircle on the blue circle traced anticlockwise.
Setting $\Theta(-\infty)=\Theta_P$ and $\Theta(\infty)=\pi+\Theta_P$ and tracing the semicircle on the blue circle anticlockwise is tantamount to the global map $\nagn\to-\nagn$ and therefore produces an equivalent wall in the antiferromagnet. 
The second one is constructed by setting $\Theta(-\infty)=\Theta_P$ and $\Theta(\infty)=\pi+\Theta_P$ and is represented by a semicircle on the red circle traced clockwise.
An equivalent wall is obtained by setting $\Theta(-\infty)=\pi+\Theta_P$ and $\Theta(\infty)=\Theta_P$ and tracing the semicircle on the red circle clockwise. 
That is, the two semicircles connecting $\Theta_P, \pi+\Theta_P$ on each one of the two circles (red and blue) represent the same domain wall.

A useful formula for the velocity is obtained if we multiply Eq.~\eqref{eq:Newton_dampingForcing} by $\Theta'$ and then integrate to find
\begin{equation} \label{eq:virial}
v \int_{-\infty}^\infty\Theta'^2\,d\xi = \frac{\storque\pi}{\alpha},
\end{equation}
which is valid for all velocities $v$ and will be used in the following. In deriving Eq.~\eqref{eq:virial}, we have assumed walls with $\Theta(\xi=-\infty)-\Theta(\xi=\infty) = \pi$ with $\Theta' < 0$ (tracing the blue circle anticlockwise in Fig.~\ref{fig:DWonCircle}).
For the same boundary conditions and $\Theta' > 0$, one obtains a wall with the opposite velocity $v\to -v$.

The magnetization is given by Eq.~\eqref{eq:totalMagnetization} for $\dot{\nagn}=-v\nagn'$.
Using the spherical parametrization \eqref{eq:sphericalParametrization} and setting $\Phi=0$, we obtain 
\begin{equation*} 
    m_1 = 0,\quad m_2 =-\frac{\epsilon v}{2\sqrt{2}}\Theta',\quad m_3 = 0.
\end{equation*}
The total magnetization for a propagating domain wall is
\[ 
\tmagn_2=-\frac{s v}{\sqrt{2}}\int_{-\infty}^{\infty}\Theta'\,d\xi
=\frac{s\pi}{\sqrt{2}}\,v.
\]
It is in-plane and remains finite for all velocities.
A faster wall not only has a higher total moment but, as we shall see in Sec.~\ref{sec:highCurrents}, this is concentrated in a narrower wall width.

\subsection{Low currents}

For low currents, systematic and detailed results can be obtained by applying a perturbation method.
The method is central for the results and arguments in this subsection, but it is relegated to an appendix so that the flow of the text is not interrupted.
We closely follow the method developed in Ref.~\cite{2021_PhysD_KomineasMelcherVenakides} and we give the results in the following.

We assume the forms
\begin{equation} \label{eq:perturbativeExpansion}
\Theta = \Theta_0 + \storque \Theta_1 + \storque^2 \Theta_2 + \ldots,\qquad v = \storque v_1 + \storque^2 v_2 + \ldots,\qquad \storque \ll 1.
\end{equation}
Substituting the series into Eq.~\eqref{eq:Newton_dampingForcing}, we obtain a sequence of linear non-homogeneous equations \eqref{eq:Theta_n} for $\Theta_n,\; n=1,2,3,\ldots$ and the corresponding solvability conditions \eqref{eq:solvabilityCondition} which determine the values of $v_n$.

In the order $O(1)$, it is obtained $\Theta_0'' = \frac{1}{2}\sin(2\Theta_0) \Rightarrow \Theta_0' = -\sin\Theta_0$, that corresponds to a standard domain wall.
In the order $O(\storque)$, we find that $\Theta_1(\xi)$ is an even function of $\xi$, according to formula~\eqref{eq:Theta_n}.
It satisfies $\Theta_1(0) = 0$, i.e., the value $\Theta=\pi/2$ remains at $\xi=0$.
Since $\Theta_0$ is odd around the value $\frac{\pi}{2}$, we see that the domain wall profile does not have a definite parity when $\storque\neq 0$.

\begin{figure}[t]
    \centering
\includegraphics[width=7cm]{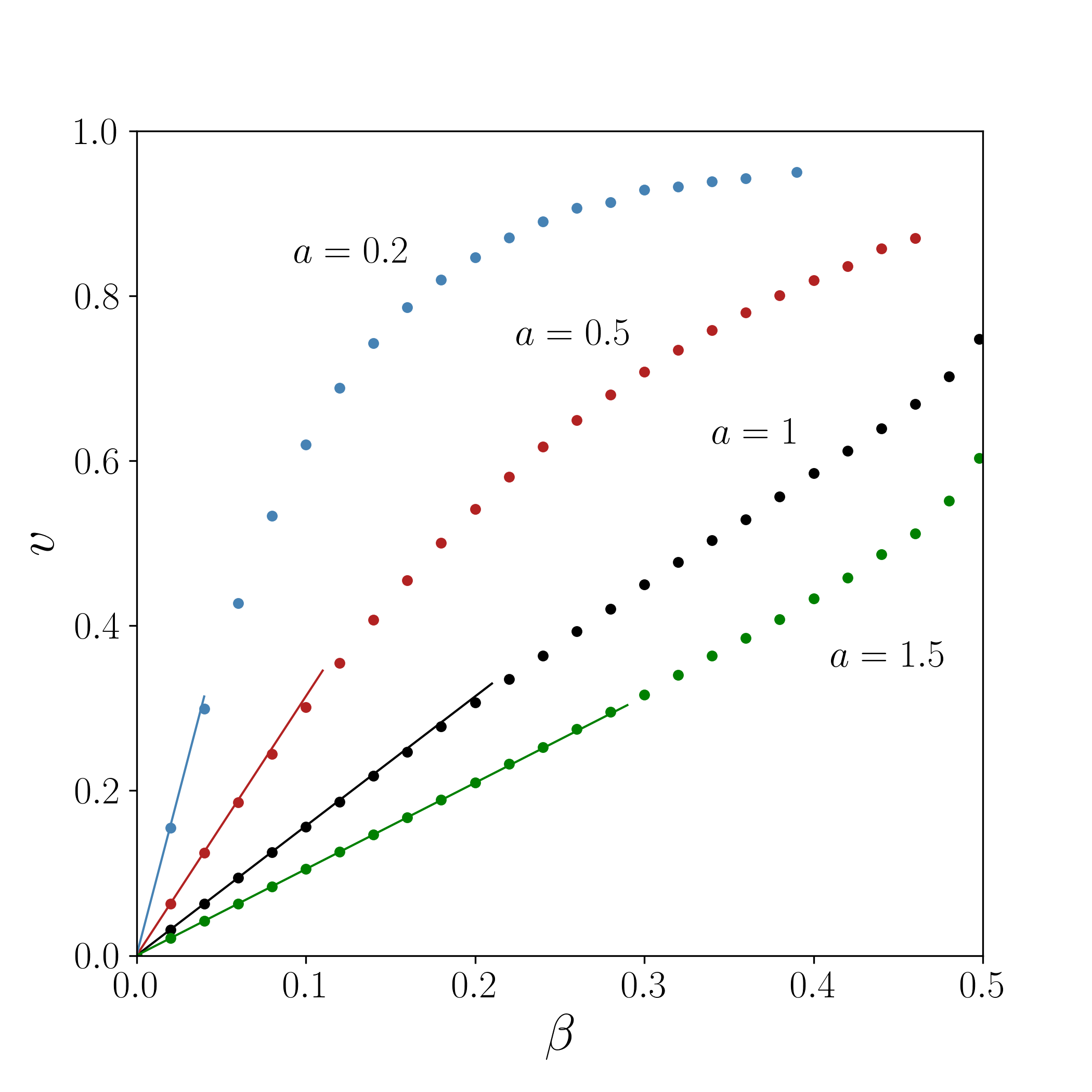}
    \caption{The velocity $v$ of the propagating domain wall as a function of the parameter $\storque$ for various values of the damping parameter $\alpha$. The spin current is polarized in $\e_2$.
    The solid lines give the theoretical formula for low currents, $v=\frac{\pi}{2}\,\frac{\storque}{\alpha}$.
    The velocity in physical units is $v_{\rm ph} = 2\sqrt{2}asJ\,v$.
    }
    \label{fig:velocity-storque}
\end{figure}

The solvability condition \eqref{eq:solvabilityCondition} gives $\alpha v_1 = \frac{\pi}{2}$ and thus, for low currents, the velocity is
\begin{equation} \label{eq:velocity_smallcurrent}
v = \frac{\pi}{2}\,\frac{\storque}{\alpha},\qquad \text{for}\qquad v\ll1.
\end{equation}
Restoring the parameters of the discrete spin chain, we have the velocity in physical units $v=\frac{\pi}{2}\frac{\tilde{\storque}}{\tilde{\alpha}}a\sqrt{\frac{J}{\Anisotropy}}$.
The approximation \eqref{eq:velocity_smallcurrent} for the velocity was obtained in Refs.~\cite{2016_PRL_ShiinoLee,2022_PRApplied_OvcharovGalkinaIvanov} employing a collective coordinate method.
Figure~\ref{fig:velocity-storque} shows the numerical results for the velocity of the domain wall as a function of the current $\storque$ for various values of damping.
They are compared with the linear formula \eqref{eq:velocity_smallcurrent} and are in excellent agreement for small $\storque$.
Further terms in the expansion are obtained in the Appendix, where we find $v_2=0$ and Eq.~\eqref{eq:v3} gives the $O(\storque^3)$ term.

Equivalent results for the expansion of the velocity can be obtained by utilizing Eq.~\eqref{eq:virial} where we insert the perturbative expansion \eqref{eq:perturbativeExpansion}.
We have
\[
\int \Theta'^2\,d\xi = \int \Theta_0'^2\,d\xi + 2\storque \int \Theta_0' \Theta_1'\,d\xi + \storque^2 \int \left(\Theta_1'^2 + 2 \Theta_0'\Theta_2' \right)\,d\xi + O(\storque^3).
\]
The first term on the right side gives $\int_{-\infty}^\infty \Theta_0'^2 d\xi = -\int_\pi^0 \sin\Theta_0\,d\Theta_0 = 2$.
The second term vanishes as $\Theta_0'$ is even and $\Theta_1'$ is odd.
We find for the velocity,
\begin{equation} \label{eq:velocity_smallcurrent_virial}
v \left[ 2 + \storque^2 \int \left(\Theta_1'^2 + 2 \Theta_0'\Theta_2' \right)\,d\xi + O\left(\storque^3\right) \right] = \frac{\storque\pi}{\alpha}
\Rightarrow v = \frac{\pi}{2}\frac{\storque}{\alpha} \left[ 1 - \frac{\storque^2}{2} \int \left(\Theta_1'^2 + 2 \Theta_0'\Theta_2' \right)\,d\xi \right] + O\left(\storque^4\right).
\end{equation}
The result agrees with \eqref{eq:velocity_smallcurrent} and with the form of the expansion for the velocity found in the Appendix.
A calculation of the precise coefficient of the $\storque^3$ term requires the explicit calculation of $\Theta_1, \Theta_2$ and will not be carried out here.
In Eq.~\eqref{eq:velocity_smallcurrent_virial}, the dependence of the integral on the parameter $\alpha$ can be found based on the properties of $\Theta_n$ discussed in the Appendix.
The first term in the integrand does not depend on any parameters, while the second also contains a term free of parameters and a term with coefficient $1/\alpha^2$.
We conclude that the velocity expansion up to $O(\storque^3)$ contains terms with the following dependence on the parameters
\begin{equation} \label{eq:velocityForm_smallcurrent_virial}
v = \frac{\pi}{2}\frac{\storque}{\alpha} - \left( \frac{\storque^3}{\alpha^3} A_1 + \frac{\storque^3}{\alpha} A_2 \right) + O(\storque^4),
\end{equation}
where $A_1, A_2$ do not depend on parameters and can be written in terms of $\Theta_0, \Theta_1, \Theta_2$ (we omit the formulae and note that $A_1,\,A_2$ do not necessarily need to be positive).

Previous approaches \cite{1978_PRA_MacLaughlinScott,2022_PRApplied_OvcharovGalkinaIvanov} have assumed propagating wall profiles resulting from a Lorentz transformation for $\Theta_0$, effectively assuming $\Theta_1=0$.
In this case, one could insert the profile $\Theta(\xi) = \Theta_0(\xi/\sqrt{1-v^2})$ in the virial relation $\eqref{eq:virial}$ to obtain
\[
\frac{v}{\sqrt{1-v^2}} = \frac{\storque}{2\alpha} \Rightarrow
v = \frac{\storque/2\alpha}{\sqrt{1+(\storque/2\alpha)^2}}.
\]
This does not contain the term $O(\storque^3/\alpha)$.
Therefore, the approximation of a Lorentz-transformed profile could give a good approximation to the velocity for small damping $\alpha$ rather than for small velocities.


\subsection{High currents}
\label{sec:highCurrents}

Figure~\ref{fig:velocity-storque} shows that for high currents the velocity reaches a maximum.
We start the study by noting that the factor in the first term of Eq.~\eqref{eq:Newton_dampingForcing} suggests a possible maximum velocity $v=1$.
For small values of $\alpha$, the current parameter $\storque$ can be varied up to a maximum value that is less than $1/2$ and we see that the velocity gets close to $v=1$ but it does not reach this value.
Above this maximum value for $\storque$, our simulations do not converge to a propagating domain wall.
Instead, a nontrivial precessional motion is observed numerically over the entire system, a phenomenon that has been noted in Ref.~\cite{2022_PRApplied_OvcharovGalkinaIvanov}.
For higher values of $\alpha$, the terminal value for $\storque$ increases and it can be very close to $\storque=1/2$.
In this case, the velocity remains clearly below unity.
We see in Fig.~\ref{fig:velocity-storque} that the terminal velocity decreases as the damping parameter $\alpha$ increases.

We consider the case where $\storque$ is very close to the value $1/2$ and Eq.~\eqref{eq:polarized_tilted} gives $\Theta_P \approx \frac{\pi}{4}$.
At the ends of the wall, we linearize Eq.~\eqref{eq:Newton_dampingForcing} and get a solution
\begin{equation}
\Theta-\Theta_P \sim e^{-\alpha v\xi/(1-v^2)}.
\end{equation}
This exponential decay is valid for $\xi\to\infty$ (when $v>0$).
At the other end of the wall, $\xi\to-\infty$, the result of the linearization is not acceptable because it diverges, and we have to keep nonlinear terms.
We set $\storque=\frac{1}{2}$ and write $\Theta = \frac{5\pi}{4}+\theta$ so that Eq.~\eqref{eq:Newton_dampingForcing} gives
\[
(1-v^2) \theta'' + \alpha v \theta' + \theta^2 = 0.
\]
The solution is $\theta=\alpha v/\xi$ (note that the first term is subleading) and the behavior of $\Theta$ is
\begin{equation} \label{eq:powerLaw}
\Theta = \frac{5\pi}{4} + \frac{\alpha v}{\xi},\qquad \xi\to -\infty.
\end{equation}
Fig.~\ref{fig:traveling_solution2} shows a domain wall profiles exhibiting asymmetry,  and this is more pronounced for $\storque=0.495$.
The wall profile decays exponentially at large $\xi$ and follows a power law for negative $\xi$.

It is useful to explore whether the velocity can reach the value $v=1$. 
Eq.~\eqref{eq:Newton_dampingForcing} would then reduce to
\begin{equation} \label{eq:v=1}
    \alpha\Theta' = \frac{1}{2} \sin(2\Theta) - \storque.
\end{equation}
For any $0 \leq \storque < \frac{1}{2}$, the right side of the equation gives that $\Theta'$ is zero not only at $\Theta_P$ and $\pi+\Theta_P$ but also for some intermediate value $\Theta_P < \Theta < \pi + \Theta_P$.
This is a fixed point of Eq.~\eqref{eq:v=1} and therefore there is no solution connecting $\Theta_P$ with $\pi+\Theta_P$. As a conclusion, there is no domain wall for $0 \leq \storque < \frac{1}{2}$ which can reach the velocity value $v=1$.
In Refs.~\cite{1974_JAP_NakajimaOnodera,1974_JAP_NakajimaYamashita}, Eq.~\eqref{eq:Newton_dampingForcing} was studied numerically as a model for magnetic fluxons \cite{1978_PRA_MacLaughlinScott}.
It is suggested that $v=1$ is obtained for $\storque<1/2$, but these numerical results in the high velocity regime do not agree with our previous analytical result.

For $\storque=\frac{1}{2}$ and velocity $v=1$, Eq.~\eqref{eq:Newton_dampingForcing} gives 
$\alpha\Theta' = \frac{1}{2} \left[\sin(2\Theta) - 1 \right]$
and has the solution
\begin{equation} \label{eq:DW_high_spintorque}
    \tan\left(\Theta + \frac{\pi}{4} \right) = -\frac{\xi-\xi_0}{\alpha}
\quad\text{or}\quad
    \tan\Theta=\frac{(\xi-\xi_0)+\alpha}{(\xi-\xi_0)-\alpha},
\end{equation}
where $\xi_0$ is an arbitrary constant.
It represents a domain wall with $\Theta = 5\pi/4$ and $\pi/4$ at the two ends of the system $x=\mp\infty$, and a width proportional to the damping parameter $\alpha$.
The wall \eqref{eq:DW_high_spintorque} appears to be a special solution of Eq.~\eqref{eq:Newton_dampingForcing}.
Nevertheless, it shares some main features with the solutions presented here: the wall profile is asymmetric around the center of the wall, and it presents a power law decay at large distances.

\begin{figure}[t]
    \centering
\includegraphics[width=7cm]{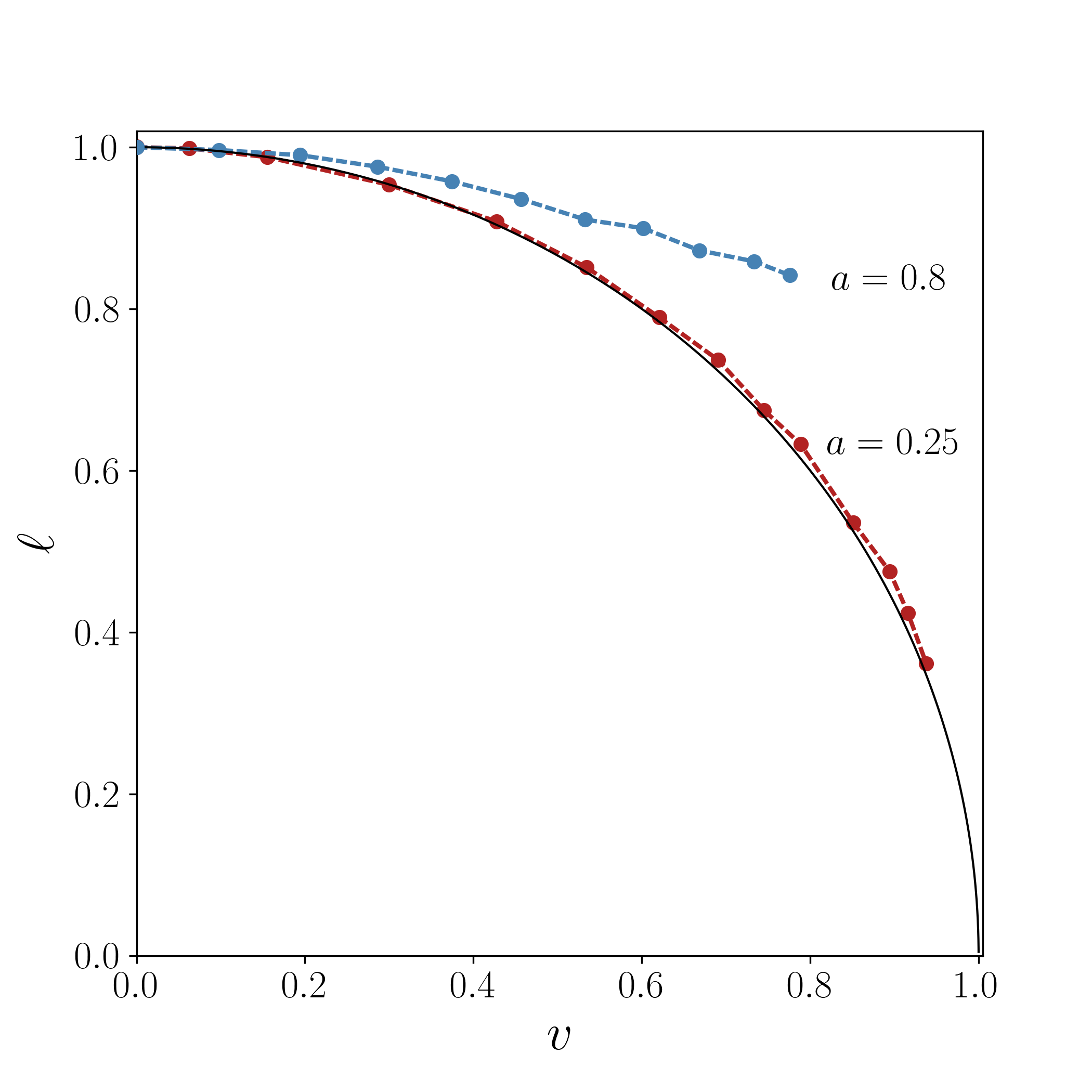}
    \caption{Propagating domain wall width, defined as $\ell=1/\Theta'$ at the point where $\Theta=\pi/2$, as a function of the velocity $v$ for damping parameter values $\alpha=0.25$ (red) and $\alpha=0.8$ (blue).
    The maximum velocity does not approach the velocity $v=1$ for either value of $\alpha$.
    The solid black line gives the domain wall width in the Lorentz transform case, $\ell=\sqrt{1-v^2}$.
    }
    \label{fig:width-velocity}
\end{figure}

We explore the dependence of the width of the domain wall on the velocity.
Given that, for larger current, the velocity $v$ increases slower than linearly with $\storque$, Eq.~\eqref{eq:virial} indicates that $\Theta'$ should become steeper as $v$ increases (so that the integral in Eq.~\eqref{eq:virial} increases).
That means that the domain wall width should decrease at higher velocities.
A possible measure of the domain wall width could be $\ell=1/\Theta'$ at the point where $\Theta=\pi/2$.
Fig.~\ref{fig:width-velocity} shows that $\ell$ decreases with $v$, but remains greater than the damping parameter $\alpha$, which is the width of the wall in Eq.~\eqref{eq:DW_high_spintorque}.
For small $\alpha$, it seems to follow the form $\ell=\sqrt{1-v^2}$ of the Lorentz transformed solution (although the maximum possible velocity does not approach unity).
This is a surprising result for two reasons. First, for larger $\storque$, the second term in Eq.~\eqref{eq:Newton_dampingForcing} that breaks the Lorentz invariance is not small.
Second, we have shown in Eq.~\eqref{eq:velocity_smallcurrent_virial} that the formula $v=v(\storque)$ for small $\storque$ does not fully agree with the formula expected from a Lorentz-transformed wall profile.
It appears that the difference is small for small damping $\alpha$.
For larger $\alpha$, we have large deviations from the Lorentz contraction formula.
It should be emphasized that $\ell$ may not be an accurate definition of the domain wall width, especially for larger $\storque$ where the wall profile is strongly asymmetric.

When an interfacial DM interaction is added, it does not modify the equation, as the corresponding term vanishes. Therefore, the result remains identical to the case without this interaction and corresponds to a propagating domain wall.

\section{Precessional dynamics}
\label{sec:precession}

Let us now consider a spin current with polarization in $\ptorque = \e_3$.
We simulate numerically the discrete Eq.~\eqref{eq:Heisenberg} for $\DM=0$ (no DM) using as an initial condition a static N\'eel type domain wall.
Eventually, the system reaches a steady state where all spins precess around $\e_3$ with constant angular frequency.
For a theoretical study, we set $\dm=0$ in the equation of motion \eqref{eq:sigmaModel_1D_spinTorque} and it gives
\begin{subequations} \label{eq:ThetaPhi_eqs}
\begin{align}
 & \ddot{\Theta}- \Theta''+ \alpha \dot{\Theta} + \left(1 + \Phi'^2 - \dot{\Phi}^2   \right) \sin \Theta\cos\Theta=0 \\
  & \sin \Theta \left(\ddot{\Phi}-\Phi'' + \alpha \dot{\Phi} - \storque \right) + 2 \cos \Theta  \left(\dot{\Theta} \dot{\Phi}-\Theta' \Phi '\right) = 0.
\end{align}
\end{subequations}
Anticipating oscillatory dynamics, we assume the simple space and time dependence,
$\Theta=\Theta(x),\;\Phi=\Phi(\tau)$
and substitute in Eqs.~\eqref{eq:ThetaPhi_eqs} to get
\begin{equation} \label{eq:ThetaPhi_eqs_oscillations}
\Theta''+\left( \dot{\Phi}^2-1\right) \sin \Theta\cos\Theta=0, \qquad
  \ddot{\Phi}+ \alpha \dot{\Phi} - \storque = 0.
\end{equation}
The second equation gives
\begin{equation} \label{eq:pressesion_frequency}
    \Phi(\tau) = \frac{c}{\alpha}e^{-\alpha \tau} + \omega \tau,\qquad \omega = \frac{\storque}{\alpha},
\end{equation}
where $c$ is a constant.
This gives a stable precession of the N\'eel vector with frequency $\omega$ for large times.
Equation~\eqref{eq:ThetaPhi_eqs_oscillations} for $\Theta$ becomes
\begin{equation} \label{eq:Theta_eq_oscillations_DW}
\Theta'' - \left( 1 - \omega^2 \right)  \sin \Theta\cos\Theta=0
\end{equation}
and it has the domain wall solution
\begin{equation} \label{eq:ThetaDW_oscillatory}
\tan\left(\frac{\Theta}{2}\right)=e^{-(x-x_0)/\losc},\qquad \losc =  \frac{1}{\sqrt{1 -  \omega^2}}
\end{equation} where $x_0$ is the position of the wall center.
This type of solution was given in \cite{1979_SJLTP_BaryakhtarIvanov,1983_SPJETP_BaryakhtarIvanov,1983_PRL_Haldane} for the conservative case, $\alpha=0,\,\storque=0$ and was further discussed in \cite{1995_PRL_IvanovKolezhuk,2017_LTP_GalkinaOvcharovIvanov}. 
The precessing domain wall is obtained here as a stable state which the system spontaneously reaches under spin torque.
The precessing state under current has been previously reported in Ref.~\cite{2022_PRApplied_OvcharovGalkinaIvanov} within a collective coordinate approach and studied as a spin-Hall nano-oscillator, although its stability was only assessed numerically.
The solution \eqref{eq:ThetaDW_oscillatory} exists for
\begin{equation}
     \omega^2 < 1 \Rightarrow -1< \frac{\storque}{\alpha} < 1.
\end{equation}

The forms \eqref{eq:pressesion_frequency}, \eqref{eq:ThetaDW_oscillatory} give a domain wall with the N\'eel vector precessing around the easy-axis $z$.
It should be noted that the precessional motion of the in-plane components of $\nagn$ can only be obtained in the presence of a domain wall.
While Eq.~\eqref{eq:Theta_eq_oscillations_DW} gives polarized solutions ($\Theta=0,\pi$) as well as a domain wall solution, it is only in the latter case that the precession around the $z$ axis makes sense.
Eq.~\eqref{eq:Theta_eq_oscillations_DW} also has the solutions $\Theta=\pm\pi/2$, but they give polarization perpendicular to the easy axis and are thus unstable.

Within the conservative Lorentz invariant model, propagating walls are obtained by boosting the stationary wall \eqref{eq:ThetaDW_oscillatory} \cite{1983_SPJETP_BaryakhtarIvanov,1983_PRL_Haldane}.
But, Lorentz invariance is broken in our case due to the damping term and propagating solutions of precessing domain walls do not exist.

We finally note that when a DM term is added in the model, no precessional motion is possible, since the DM interaction breaks the symmetry of rotations around the $z$ axis.
In this case, a static domain wall is obtained that is of mixed Bloch-N\'eel type.

For the perpendicular component of the total magnetization, we find
\begin{equation}
\tmagn_3 = \frac{4s}{2\epsilon} \int_{-\infty}^\infty m_3\,dx = \sqrt{2}s\omega\losc = \sqrt{2}s\frac{\omega}{\sqrt{1-\omega^2}},
\end{equation}
a result that is mentioned in \cite{1983_PRL_Haldane,1995_PRL_IvanovKolezhuk}.
Since $m_1, m_2$ are odd functions, the in-plane total magnetization is zero.
However, it is useful to calculate the in-plane magnetization on the left and the right side of the wall.
For the right side, it is
\[
    \tmagn_1 + i \tmagn_2 = \frac{4s}{2\epsilon} \int_0^\infty (m_1+i m_2)\,dx = -\frac{s}{\sqrt{2}}\frac{\omega}{\sqrt{1-\omega^2}}\,e^{i\omega \tau}
\]
and the opposite result is found for the left side.
The magnetization diverges to infinity for precession frequencies $\omega\to 1$ as a result of the domain wall width $\ell$ in Eq.~\eqref{eq:ThetaDW_oscillatory} diverging in the same limit.
The significant increase of the magnetization can provide a method for the detection of antiferromagnetic domain walls by either direct measurements of the magnetization or by its possible effect in adjacent magnets.

\section{Domain wall oscillations}
\label{sec:oscillations}

We now choose a current with $\storque\ptorque = \storque_2\e_2 + \storque_3\e_3$.
We consider an initially static N\'eel-type domain wall and simulate numerically the spin equations \eqref{eq:Heisenberg}.
The simulations show that, after a transient period, a domain wall undergoes oscillatory motion.
The wall is seen to move periodically between two end positions, $x_{\rm min} < x < x_{\rm max}$.

For studying the oscillating motion of the domain wall, we write the equations for $\Theta, \Phi$,
\begin{subequations} \label{eq:ThetaPhi_p2_p3}
\begin{align}
 & \ddot{\Theta} - \Theta''+ \alpha \dot{\Theta} + \left(1 + \Phi'^2 - \dot{\Phi}^2\right) \sin \Theta\cos\Theta = \storque_2\cos\Phi  \label{eq:Theta_p2_p3} \\
  & \sin \Theta \left(\ddot{\Phi} - \Phi'' + \alpha \dot{\Phi} - \storque_3 \right)+2 \cos \Theta \left(\dot{\Theta} \dot{\Phi}-\Theta' \Phi' \right) = -\storque_2\cos\Theta\sin\Phi.  \label{eq:Phi_p2_p3}
\end{align}
\end{subequations}
Polarized states of Eqs.~\eqref{eq:ThetaPhi_p2_p3} satisfy
\begin{equation} \label{eq:oscillating_polarized}
\storque_2 \cos\Phi = \frac{1}{2}\sin(2\Theta),\qquad
\storque_2 \sin\Phi = \storque_3\tan\Theta.
\end{equation}
Eliminating $\Phi$ and using the identity $\sin(2\Theta) = 2\tan\Theta/(1+\tan^2\Theta)$ obtains
\begin{equation} \label{eq:oscillating_polarized_theta}
    \storque_3^2\tan^2\Theta + \frac{\tan^2\Theta}{(1+\tan^2\Theta)^2} = \storque_2^2.
\end{equation}
This gives $\tan^2\Theta$ as the root of a third order polynomial.
It has a single root, at least for reasonable values of $\storque_2, \storque_3$.
As a result, we have two roots for the angle, $\Theta=\Theta_P$ and $\Theta=\pi-\Theta_P$.
Eqs.~\eqref{eq:oscillating_polarized} give for the angle $\Phi$ two corresponding values denoted $\Phi_P$ and $\pi+\Phi_P$.
The states $(\Theta_P,\Phi_P)$ and $(\pi-\Theta_P, \pi+\Phi_P)$ are represented on the Bloch sphere by two antipodal points.

\begin{figure}[t]
    \centering
    \subfloat[]
{\includegraphics[width=6.0cm]{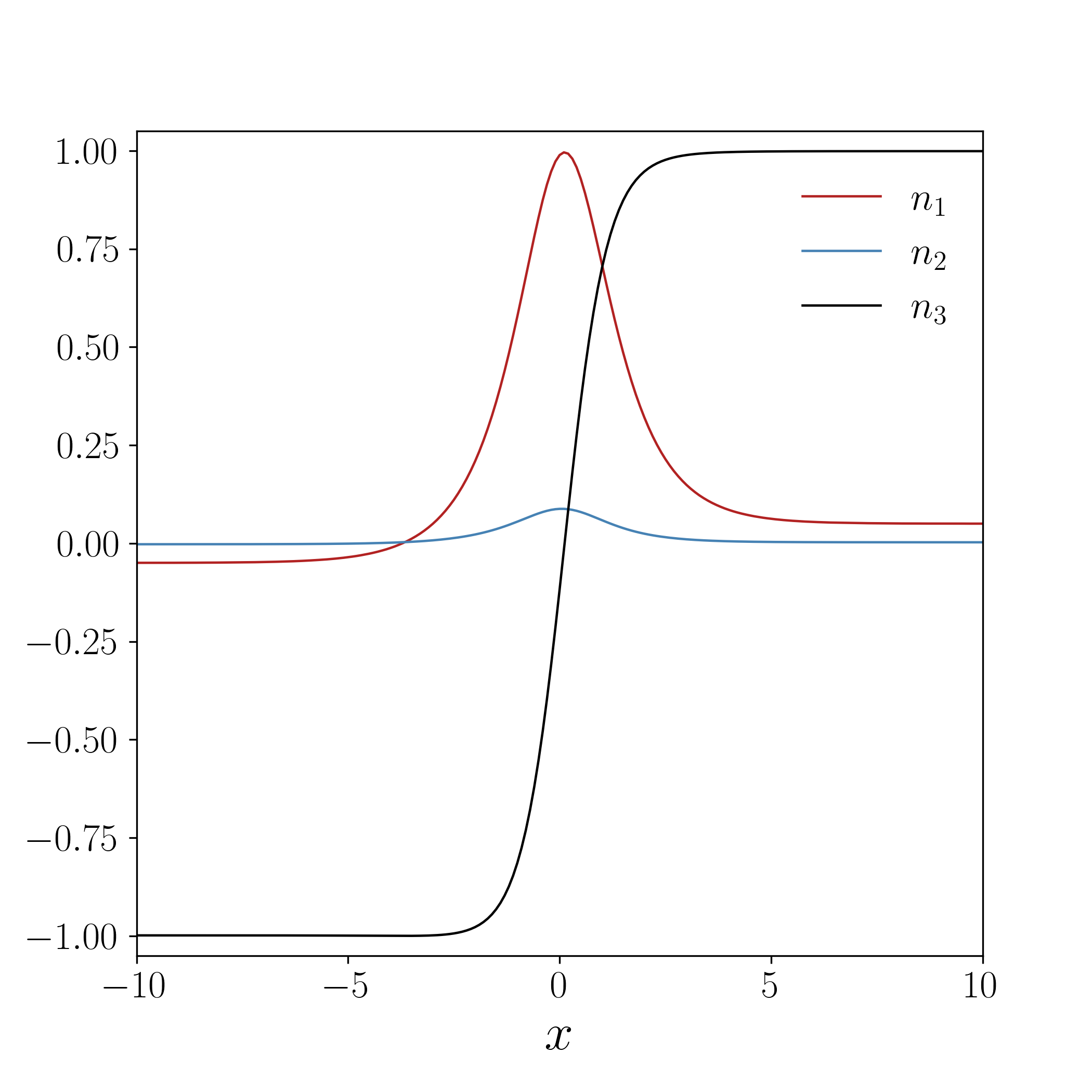}}\qquad
   \subfloat[]
{\includegraphics[width=6.0cm]{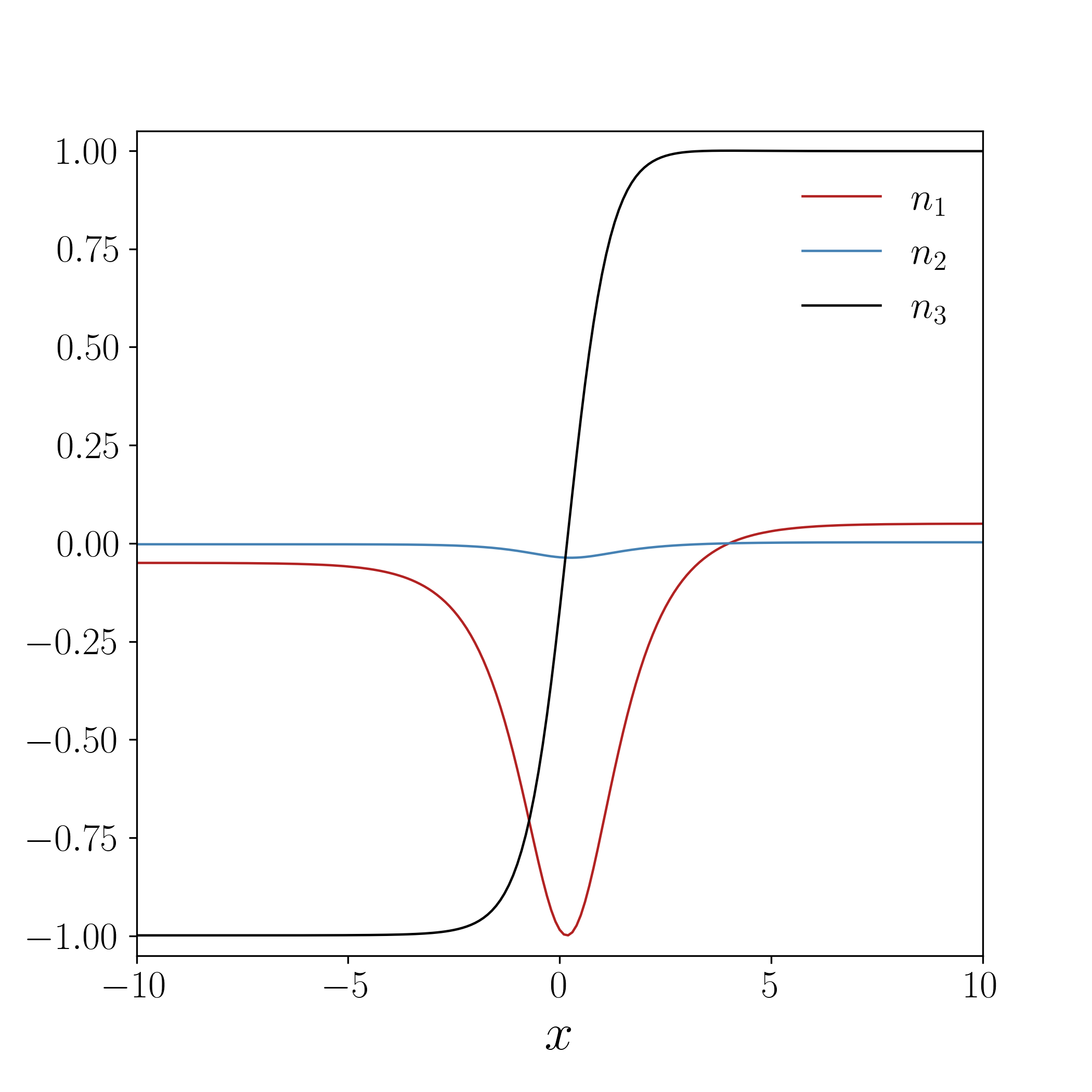}}
    \caption{Under spin current with $\storque\ptorque = \storque_2\e_2 + \storque_3\e_3$, the domain wall moves periodically between two end positions, $x_{\rm min} < x < x_{\rm max}$. 
    We use $\storque_2=\storque_3=0.05$ and damping $\alpha=1$.
    Two snapshots of the motion are show where the domain wall is moving (a) to the right and (b) to the left.
    In both entries, the domain wall is shown at approximately the central position, where the velocity is maximum.
    Only the central part of the numerical mesh is shown.
}
\label{fig:oscillatory_dw_snapshots-1}
\end{figure}

We expect that propagation of the domain wall will be induced by $\storque_2$ as discussed in Sec.~\eqref{sec:propagatingDW}.
Also, $\storque_3$ will induce precession of the N\'eel vector with angular frequency $\omega=\storque_3/\alpha$ as discussed in Sec.~\ref{sec:precession}, Eq.~\eqref{eq:pressesion_frequency}. However, since the ends of the wall have to be pinned to the values set by Eqs.~\eqref{eq:oscillating_polarized}, precession may only happen in the central region. Figure~\ref{fig:oscillatory_dw_snapshots-1} shows domain wall oscillations for $\storque_2=0.05,\,\storque_3=0.05$.
While the wall position oscillates, $\nagn$ is precessing around the easy axis, as seen by the change of its in-plane component.
On the other hand, the ends of the wall are seen to be pinned and $\nagn$ does not precess there.

\begin{figure}[t]
    \centering
\subfloat[]
{\includegraphics[width=6.0cm]{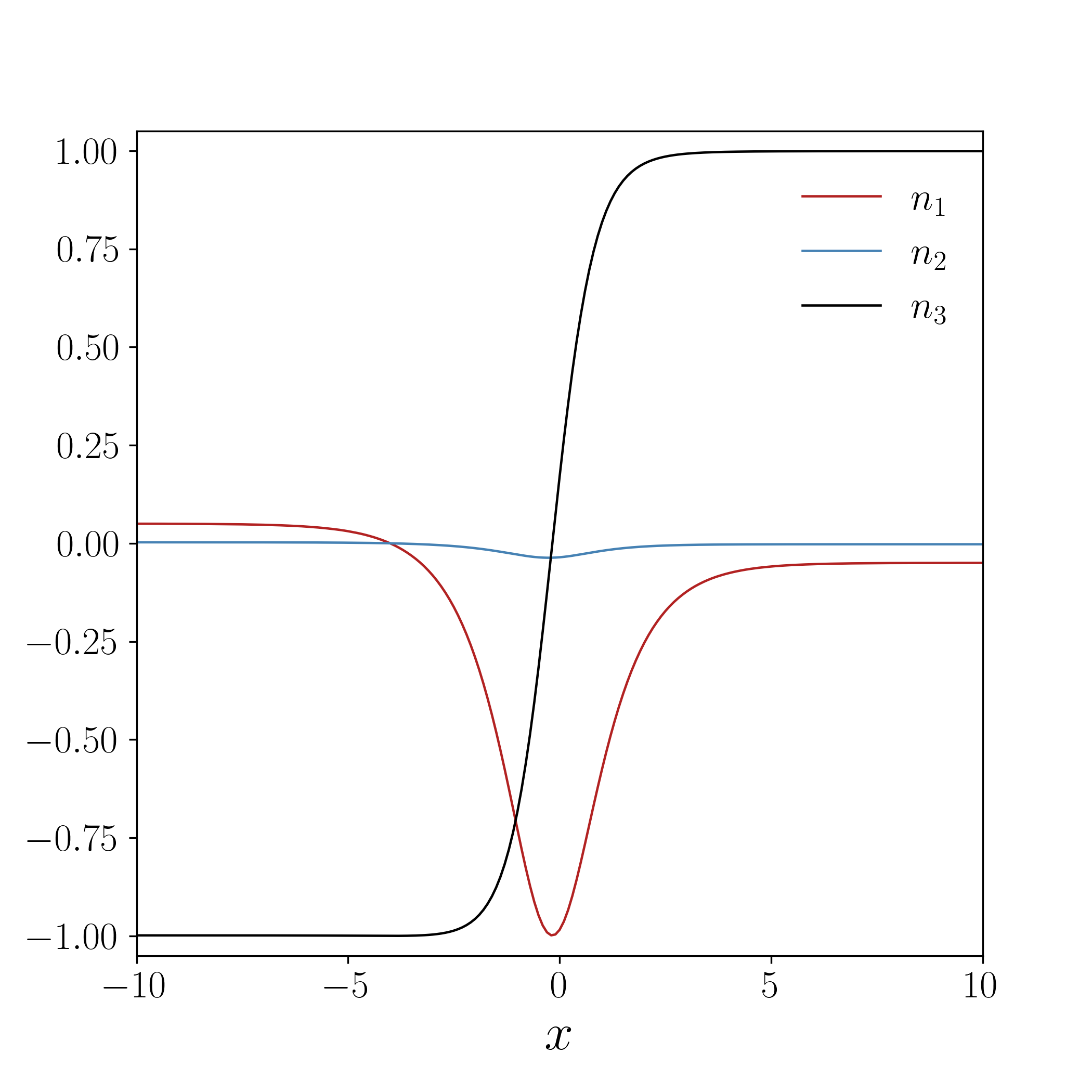}}
\qquad
\subfloat[]
{\includegraphics[width=6.0cm]{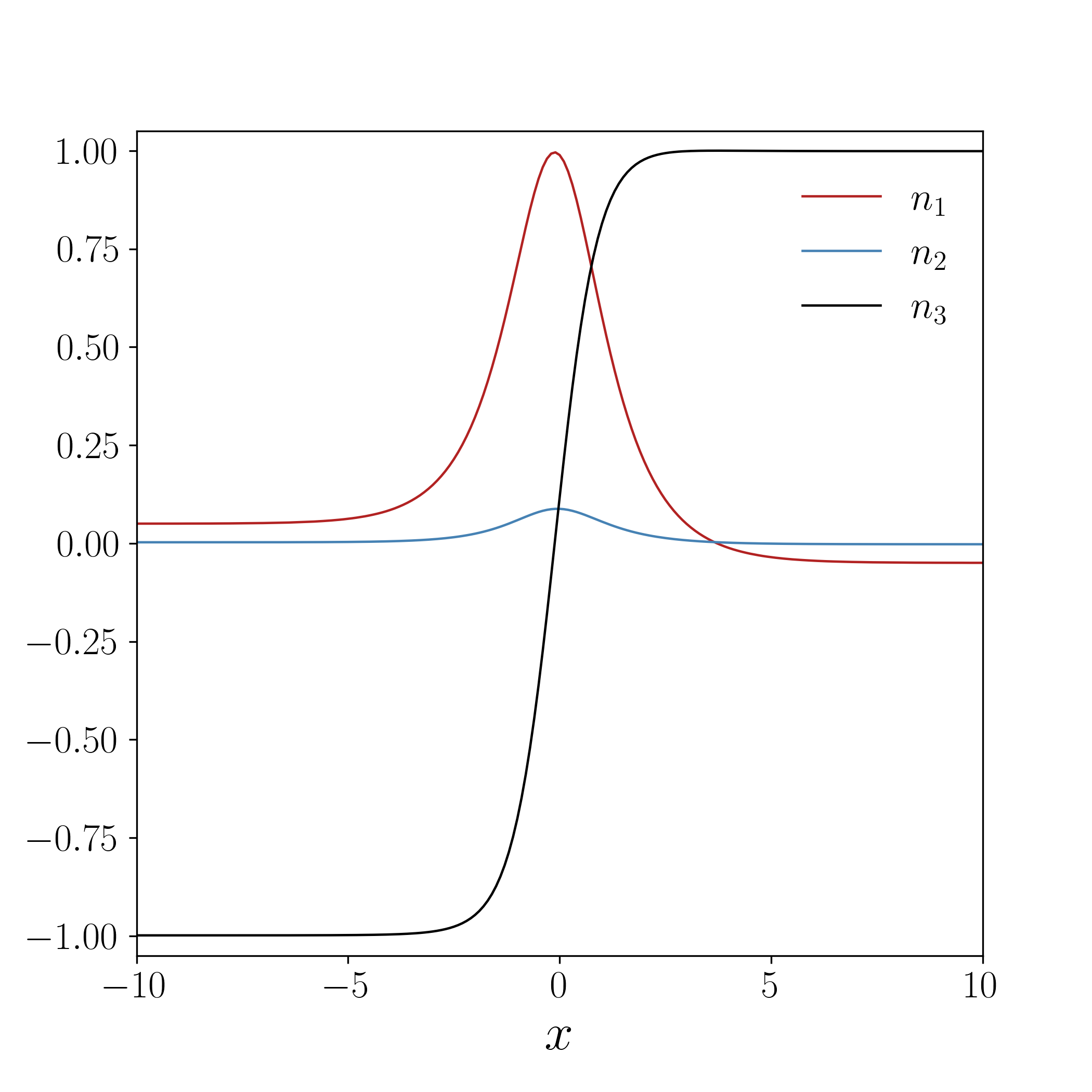}}
    \caption{Snapshots of a simulation for a system as in Fig.~\ref{fig:oscillatory_dw_snapshots-1} but with $\storque_2=-0.05,\; \storque_3=0.05$.
    Snapshots for the domain wall while it is moving  (a) to the right and (b) to the left.
    While the motion is periodic as in Fig.~\ref{fig:oscillatory_dw_snapshots-1}, the domain wall configuration is different.
}
\label{fig:oscillatory_dw_snapshots-2}
\end{figure}

Changing the sign of $\storque_2\to-\storque_2$, the dynamical equations \eqref{eq:ThetaPhi_p2_p3} give that the solution will transform according to $\Theta\to\Theta$ and $\Phi\to\Phi+\pi$.
Figure~\ref{fig:oscillatory_dw_snapshots-2} shows wall oscillations for oppositely signed $\storque_2=-0.05$ and the same $\storque_3=0.05$ compared to Fig.~\ref{fig:oscillatory_dw_snapshots-1}.
The ends of the wall are seen to be pinned to values that are rotated by $\pi$ in $\Phi$ compared to Fig.~\ref{fig:oscillatory_dw_snapshots-1}, in agreement with Eq.~\eqref{eq:oscillating_polarized}.
On the other hand, the central part of the wall is precessing with the same frequency as in Fig.~\ref{fig:oscillatory_dw_snapshots-1}.
Therefore, a domain wall undergoes oscillations with the same frequency for either positive or negative $\storque_2$, but each sign of $\storque_2$ gives the wall configuration a different flavor. 

\begin{figure}[t]
   \centering
   \subfloat[]{\includegraphics[width=6cm]{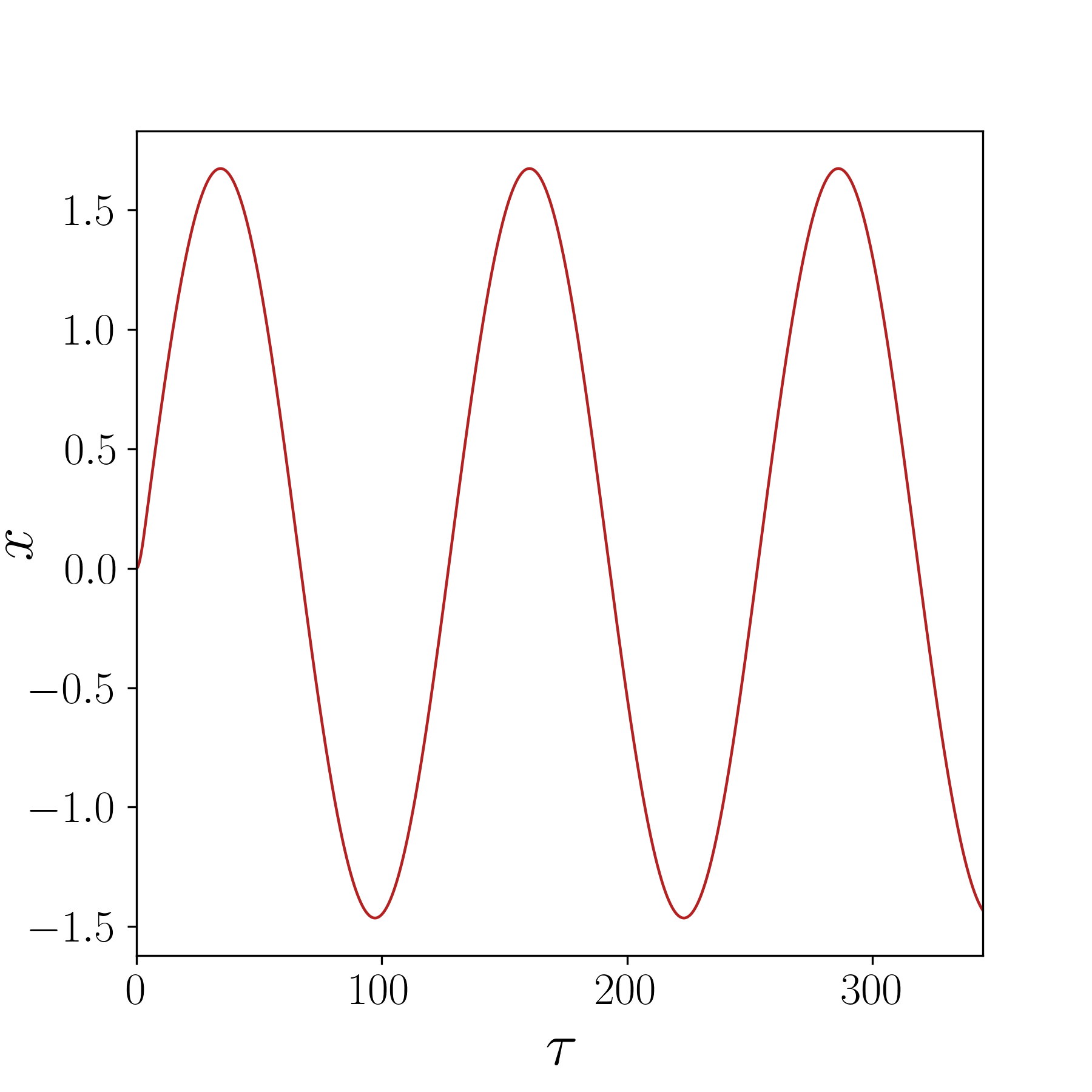}}\qquad
   \subfloat[]{\includegraphics[width=6cm]{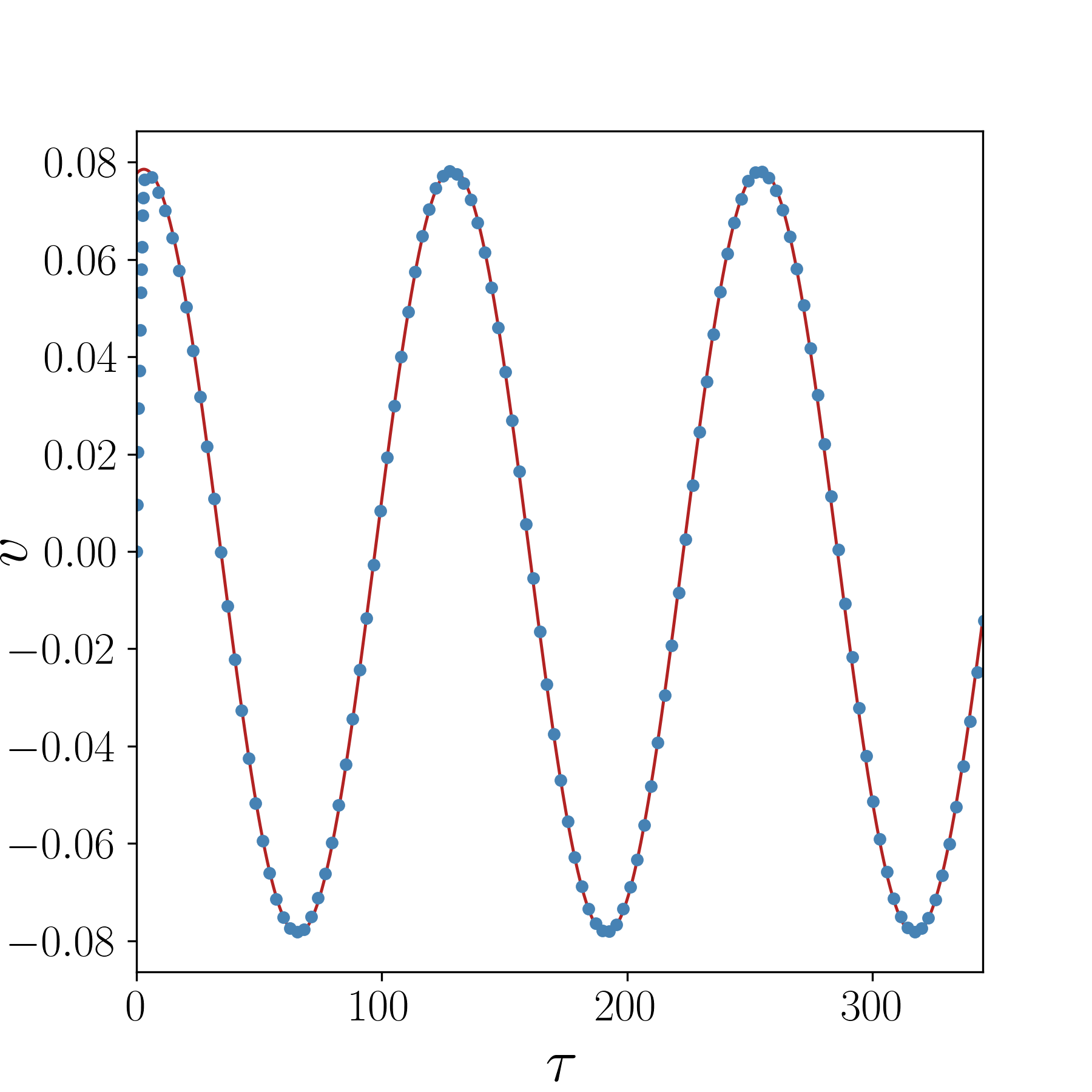}}
\caption{(a) The center $x$ (point where $\Theta=\pi/2$) of an oscillating domain wall as a function of time given by a simulation for parameter values $\storque_2=\storque_3=0.05$ and $a=1$. 
(b) The velocity $v$ in the simulation is shown by blue points.
This is compared with the result \eqref{eq:velocity_oscillations} shown by a red solid line.
The current is applied at $\tau=0$ and the system very rapidly sets in the oscillating steady state.}
\label{fig:oscillatory_dw_position_velocity}
\end{figure}

The main features of the motion can be obtained by focusing on the central region and assuming $\Phi=\omega \tau$ and $\Theta=\Theta(x-v\tau)$. Eq.~\eqref{eq:Theta_p2_p3} becomes
\begin{equation} \label{eq:oscillating_force}
    (1-v^2)\Theta'' + \alpha v \Theta' - \left(1 - \omega^2 \right) \sin \Theta\cos\Theta = -\storque_2\cos(\omega \tau).
\end{equation}
This is similar in form to Eq.~\eqref{eq:Newton_dampingForcing} which was studied in Sec.~\ref{sec:propagatingDW}.
We may tentatively use the result in Eq.~\eqref{eq:velocity_smallcurrent}, which now takes the form
\begin{equation} \label{eq:velocity_oscillations}
    v = \frac{\pi}{2} \frac{\storque_2}{\alpha}\,\cos\left(\frac{\storque_3}{\alpha} \tau \right)
\end{equation}
and gives periodic in space oscillations of the domain wall. Figure~\ref{fig:oscillatory_dw_position_velocity} shows the position and the velocity of the domain wall for the simulation in Fig.~\ref{fig:oscillatory_dw_snapshots-1}.
Figure~\ref{fig:oscillatory_dw_position_velocity}b also shows a comparison between the velocity of the domain wall in the simulation and the form \eqref{eq:velocity_oscillations}.
The agreement is excellent with respect to both the amplitude and the frequency of oscillations. Figure~\ref{fig:oscillatory_dw_position_velocity} shows that the oscillatory state is very rapidly achieved, and this is due to the large damping ($\alpha=1$) used in the simulation.
The result \eqref{eq:velocity_oscillations} is expected to be valid for low velocities, $|v| \ll 1$, since it is based on the perturbative method of Sec.~\ref{sec:propagatingDW}.
Also, we have ignored Eq.~\eqref{eq:Phi_p2_p3}, which should be a valid approximation for $|\omega| \ll 1$.
In conclusion, the result \eqref{eq:velocity_oscillations} is expected to be valid for $\storque_2, \storque_3 \ll 1$, which also imply $|v|, |\omega| \ll1$. 
Nevertheless, the accuracy of the result \eqref{eq:velocity_oscillations} for the simulation in Fig.~\ref{fig:oscillatory_dw_position_velocity} is surprising given the approximations involved in its derivation.

When the DM interaction is added, the precessional motion does not occur as already mentioned in Sec.~\ref{sec:precession}.
In the present case, the lack of precessional motion means that oscillating motion is not expected to occur either.
Numerical simulations show that a domain wall under polarization $\storque_2\e_2 + \storque_3 e_3$ for small $\storque_3$, propagates with a constant velocity, and no oscillating motion occurs.

\section{Concluding remarks}
\label{sec:conclusions}

We have studied the dynamics of antiferromagnetic domain walls under spin-orbit torques.
All simulations have been performed using the spin equations \eqref{eq:Heisenberg}, while the theoretical analysis is based on the continuum model \eqref{eq:sigmaModel_1D_spinTorque}.
This is an extension of the $\sigma$-model with the addition of damping and source terms.
The present study obtains results beyond those inferred by the idealized $\sigma$-model.

We find that the system spontaneously converges to a stationary state and the result depends on the spin polarization of the injected current.
Propagation of a domain wall with a constant velocity is obtained for in-plane spin current polarization perpendicular to the in-plane component of the wall.
The details of the dynamics are not drawn from a Lorentz transformation of the static wall, unlike in the conservative case.
Instead, perturbation and asymptotic methods are applied and give an asymmetric wall profile with a power-law tail.
Precessional dynamics, analogous to that discussed in early papers for the conservative model, is obtained as a stable steady state for perpendicular spin current polarization.
Periodic motion of the wall between two positions is obtained for a current polarization with both an in-plane and a perpendicular component. This comes in two flavors for the two different signs of the in-plane component of the current polarization.

When the Dzyloshinksii-Moriya interaction is present, we find significant modifications of the dynamical behavior.
Propagation is still found but precessional motion is not possible.
As a result, oscillating motion is not possible either, and propagation is seen also when current polarization is not fully in-plane, at least for small currents.

We discuss the magnetic moment carried by dynamical antiferromagnetic domain walls.
This depends on the wall dynamics and may also depend on the wall profile. 
For example, propagating walls have a magnetization proportional to the velocity and their width decreases with increasing velocity. As a result, faster walls have a higher local moment density and a higher total moment. However, this is bounded by the moment of a single spin. By contrast, precessing walls can have an arbitrarily large total moment for high precession frequencies.
The net moment in antiferromagnets could give a handle for the observation and manipulation of AFM textures.

Given the recent progress in methods to observe AFM order, we hope that our results will motivate detailed observations of domain wall dynamics and will help develop processes for transmission of information and logic gate implementations.

\acknowledgments
G. Theodorou gratefully acknowledges support from the ERC starting grant 101078061 SINGinGR, under the European Union's Horizon Europe program for research and innovation.
This work has greatly benefited from networking activities of the COST Action Polytopo, CA23134, supported by COST (European Cooperation for Science and Technology).

\appendix

\section{Wall profile and velocity by perturbative expansion}

For small currents, $\storque \ll 1$, we assume a perturbative expansion for the wall profile
\[
\Theta = \Theta_0 + \storque \Theta_1 + \storque^2 \Theta_2 + \ldots
\]
and for its velocity
\[
v = \storque v_1 + \storque^2 v_2 + \storque^3 v_3 + \ldots
\]
We substitute the series in Eq.~\eqref{eq:Newton_dampingForcing} and obtain a sequence of equations for successive orders of $\storque$.
To zeroth order, we obtain $\Theta_0'' = \frac{1}{2}\sin(2\Theta_0) \Rightarrow \Theta_0' = -\sin\Theta_0$ that gives the standard domain wall.
Higher orders $O(\storque^n)$ give
\begin{equation} \label{eq:Theta_n}
\Theta_n'' - \cos(2\Theta_0) \Theta_n = g_n,\qquad n=1,2,\ldots.
\end{equation}
The non-homogeneous parts for the first two equations in the sequence are
\begin{align*}
g_1 & = -1 - \alpha v_1 \Theta_0' \\
g_2 & = v_1^2 \Theta_0'' - \alpha v_2 \Theta_0' - \alpha v_1 \Theta_1' - \sin(2\Theta_0) \Theta_1^2.
\end{align*}

We will follow a method developed in Ref.~\cite{2021_PhysD_KomineasMelcherVenakides} to find $\Theta_n$ and $v_n$.
The homogeneous part of Eq.~\eqref{eq:Theta_n} has the explicit basis solutions
\[
    H_1=\sech(\xi), \qquad H_2=\sinh(\xi)+\xi\sech(\xi).
\]
Using the formula of the variation of constants (with Wronskian $W=2$), we obtain
\begin{equation} \label{eq:Theta_n_variation-constants}
  \Theta_n(\xi) = -\frac{1}{2}H_1(\xi)\int_{0}^{\xi} g_n H_2(\tau)\,d\tau + \frac{1}{2}H_2(\xi)\int_{-\infty}^{\xi} g_n H_1(\tau)\,d\tau.
\end{equation}
Satisfying the boundary conditions imposes the solvability condition
\begin{equation} \label{eq:solvabilityCondition}
\int_{-\infty}^\infty g_n H_1(\xi)\, d\xi = 0.
\end{equation}
Applying the solvability condition to $g_1$, we obtain
\begin{equation} \label{eq:v1}
    \int_{-\infty}^{\infty} [1 + \alpha v_1 \Theta_0'(\tau)] H_1(\tau)\,d\tau=0
    \Rightarrow v_1 = \frac{\pi}{2\alpha},
\end{equation}
where we used $\Theta_{0}'= -\sech\xi$.
We notice that $g_1$ is even and therefore formula \eqref{eq:Theta_n_variation-constants} gives that $\Theta_1$ is an even function. Also, since $g_1$ does not contain any parameters, $\Theta_1$ does not depend on any parameters either.

Moving to $O(\storque^2)$, we notice that all terms in $g_2$ are odd except the one that contains $v_2$.
The solvability condition \eqref{eq:solvabilityCondition} gives
\[
v_2 = 0.
\]
Since all the remaining terms in $g_2$ are odd, we have that $\Theta_2$ is an odd function.
Inspecting $g_2$, we see that the first term has a coefficient $1/\alpha^2$ and the two last terms do not contain any of the parameters.
These will produce terms in $\Theta_2$ with corresponding dependences on the parameters.

We now go to $O(\storque^3)$ and have
\begin{equation}
    g_3 = v_1^2\Theta_1'' - \alpha v_3 \Theta_0' - \alpha v_1 \Theta_2' - 2\sin(2\Theta_0)\Theta_1\Theta_2 - \frac{2}{3}\cos(2\Theta_0) \Theta_1^3.
\end{equation}
All terms on the right side in $g_3$ are even, therefore $\Theta_3$ is even.
The domain wall profile will be asymmetric.
The solvability condition gives
\begin{equation} \label{eq:v3}
v_3 = \frac{1}{2\alpha} \int_{-\infty}^\infty \left[ v_1^2\Theta_1'' - \alpha v_1 \Theta_2' - 2\sin(2\Theta_0)\Theta_1\Theta_2 - \frac{2}{3}\cos(2\Theta_0) \Theta_1^3 \right] \sech\xi\, d\xi.
\end{equation}

\bigskip
\bibliography{references.bib}
\bigskip

\end{document}